\documentclass[aps,prb,amsmath,amssymb,reprint,longbibliography,superscriptaddress]{revtex4-2}

\usepackage{graphicx}
\usepackage{dcolumn}
\usepackage{bm}

\usepackage[utf8]{inputenc}
\usepackage[T1]{fontenc}
\usepackage{mathptmx}
\usepackage{etoolbox}
\usepackage{appendix}
\usepackage{hyperref}

\usepackage{mathtools}

\DeclarePairedDelimiterX\braket[2]{\langle}{\rangle}{#1\,\delimsize\vert\,\mathopen{}#2}

\makeatletter
\def\@email#1#2{%
 \endgroup
 \patchcmd{\titleblock@produce}
  {\frontmatter@RRAPformat}
  {\frontmatter@RRAPformat{\produce@RRAP{*#1\href{mailto:#2}{#2}}}\frontmatter@RRAPformat}
  {}{}
}%
\makeatother
\begin{document}

\preprint{AIP/123-QED}

\title[Visualization of atomistic optical waves]{Visualization of atomistic optical waves in crystals}
\author{Jungho Mun}
    \affiliation{Electrical and Computer Engineering, Purdue University, West Lafayette, Indiana 47907, USA}
    \affiliation{POSCO-POSTECH-RIST Convergence Research Center for Flat Optics and Metaphotonics,\\ Pohang University of Science and Technology (POSTECH). Pohang 37673, Republic of Korea}
\author{Sathwik Bharadwaj}
    \affiliation{Electrical and Computer Engineering, Purdue University, West Lafayette, Indiana 47907, USA}
\author{Zubin Jacob}
    \email{zjacob@purdue.edu}
    \affiliation{Electrical and Computer Engineering, Purdue University, West Lafayette, Indiana 47907, USA}
    \affiliation{Purdue Quantum Science and Engineering Institute, Birck Nanotechnology Center, Purdue University, West Lafayette, Indiana 47907, USA}

\begin{abstract}
The refractive index of a matter is foundational to quantify the light-matter interaction of the medium. However, the classical description of refractive index is based on macroscopic homogenization and is limited to describing the local optical response of materials. A complete quantum description of light-matter interaction should consider nonlocality and multiple-scattering of optical responses at the atomistic lattice level. 
Recently, the deep microscopic optical band structure was introduced as a quantum generalization of refractive index of a medium.
This quantum description unveils multiple optical eigenmodes in crystalline solids and hidden microscopic optical waves at the lattice level.
In this work, we unravel the microscopic optical waves in silicon carbide. We predict and visualize hidden microscopic optical eigenwaves, which can be nonplanar and inhomogeneous even near the optical limit. Also, the nonlocal macroscopic dielectric constant of the crystal is analyzed using the microscopic optical waves as the basis.
Our work establishes a general framework for picoscale electrodynamics applicable to other materials including two-dimensional materials.
    
\end{abstract}

\date{\today}

\maketitle


\section{Introduction}

\begin{figure}[]
    \centering
    \includegraphics[scale=0.5]{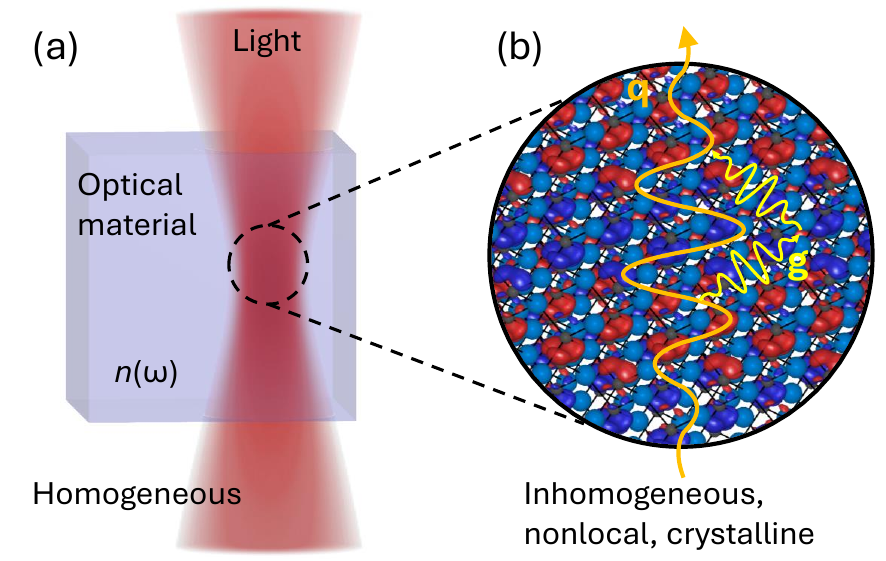}
    \caption{Concept of refractive index in a crystalline solid. (a) Classical macroscopic refractive index in the local, continuum, homogeneous limit. (b) Generalized refractive index with nonlocal, inhomogeneous optical response of atomic lattices.}
    \label{fig:concept}
\end{figure}

Refractive index has long been used to characterize the optical properties of materials~\cite{jackson1999} (Fig.~\ref{fig:concept}a). In other words, the quest of finding a novel optical material is equivalent to tuning the refractive index~\cite{andreoli2021,shim2021}. For instance, metamaterials with artificial optical properties, such as negative index~\cite{shalaev2007}, artificial-chiral~\cite{mun2020}, hyperbolic~\cite{lee2022}, epsilon-near-zero~\cite{liberal2017}, have been realized by engineering subwavelength structural compositions. This classical concept of refractive index has been successful in not only classical optics and nanophotonics, but also in quantum optics, where the spontaneous emission rates of quantum emitters have been predicted using refractive index of various electromagnetic environment~\cite{galfsky2015,gonccalves2020}. However, recent advances on photonic crystals, resonant and nonlocal meta-optics, and nanoscale optics have reached regimes with strong nonlocality or spatial dispersion where the classical homogenization and conventional effective medium theories fail~\cite{tsukerman2017}; inevitably, the concept of refractive index needs to be generalized beyond the local, continuum, and homogeneous limit~\cite{orlov2011,torquato2021,lobet2022}.

The effect of nonlocality is especially pronounced in plasmonics, which has become one of the most active research areas in nano-optics. For its extreme light localization and electric field amplification, plasmonics with nano-gaps can strongly enhance light-matter interactions at the nanoscale~\cite{gonccalves2020}. However, the length scale of plasmonic modes become highly subwavelength in such pico- or nano-structured plasmonic systems, so the classical macroscopic homogenization does not hold~\cite{mortensen2021}. As a quantum correction of dielectric constant by considering the electron spill-out at the metal surfaces, the nonlocal hydrodynamic model has been proposed ~\cite{baumberg2019,boroviks2022}. This model considers mostly surface effects and is limited to macroscopic continuum limit.

Classical description of macroscopic polarization has been related to the density of dipole moment, but this description fails to explain the polarization in periodic crystals without considering surfaces, because the dipole moment in the periodic lattice is arbitrary depending on the choice of the origin in the lattice. This discrepancy was remedied by the so-called modern theory of polarization~\cite{king1993,resta1993}, which defines macroscopic polarization through topological argument. Still, this theory does not apply to the dynamic optical refractive index of crystals.

Refractive index relates to macroscopic fields by macroscopic Maxwell's equations. This description does not consider microscopic fields that are rapidly varying at the atomic scales, where atoms or molecules in the media experience microscopic fields, or the local fields. The connection between macroscopic and microscopic quantities has been explained by the Claussius-Mossotti (Lorentz-Lorenz) equation that relates the bulk refractive index to the molecular (microscopic) polarizability~\cite{aspnes1982}, but this description is classical and only applies to isotropic or simple cubic lattices.

\begin{figure*}[]
    \centering
    \includegraphics[scale=0.5]{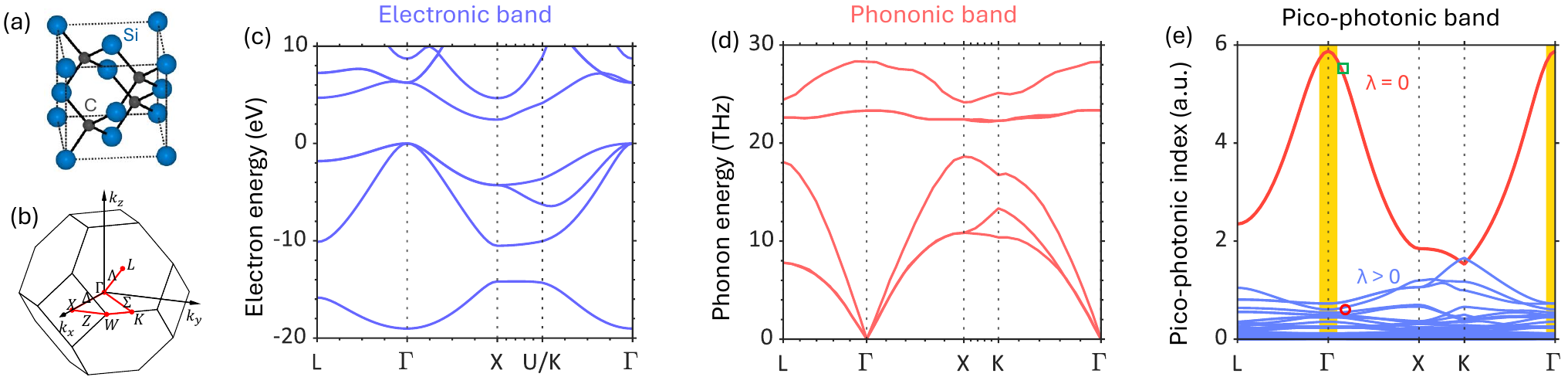}
    \caption{Light inside lattices of a crystalline solid. (a) The unit-cell and (b) the Brillouin zone of SiC with zincblende crystal structure. (c) The electronic and (d) phononic band structures of SiC. (e) The deep microscopic optical band structure $\pi_\lambda(\mathbf{q},\omega)$ of SiC. The band structure consists of the dominant band ($\lambda=0$, red solid line) and higher-order bands ($\lambda>0$, blue solid lines). The real-space distributions of some selected bands are visualized in Fig.~\ref{fig:dist}. The yellow shaded region indicates the classical optical regime with small momentum $|\mathbf{q}|$. The frequency is $\omega$ = 10~THz. The phonon band structure result was taken from Ref.~\cite{wang2017}}
\label{fig:2}
\end{figure*}

The optical properties of real crystals have also been calculated via full quantum theories, such as density functional theory, considering nonlocality, atomistic inhomogeneity or local field effect, and finite photon frequency~\cite{adler1962,wiser1963,gajdovs2006}. These calculations have allowed the numerical investigation of optical properties of crystals including optical measurements and electron energy loss spectra, but mostly limited to macroscopic properties and at the optical limit. 
Recent investigations on the atomistic electrodynamics inside crystals incorporate microscopic~\cite{mahon2019} and multiple-scattering optical responses~\cite{andreoli2023}. Along these lines, the concept of deep microscopic optical waves in crystals at finite photon momentum and the relevant software \texttt{Purdue-PicoMax} is under development~\cite{bharadwaj2024}.

In this work, we investigate the optical responses and hidden polarization waves of SiC in the atomistic nonlocal regime, where the nonlocal, inhomogeneous, and multiple-scattering optical responses are considered (Fig.~\ref{fig:concept}b). We introduce the deep microscopic optical band structure of SiC as a quantum generalization of refractive index of the medium. Contrary to the classical refractive index with single-modal nature, multiple-scattering due to atomic inhomogeneities allow multi-modal optical responses~\cite{andreoli2023}. Also, the effective optical response, or the effective screening strength, experienced by an arbitrary field inside the crystal is introduced. This effective screening is obtained by decomposing the field into a set of deep microscopic optical eigenwaves with distinct screening strengths. Our framework provides the fundamental understanding of the refractive index and the light-matter interaction inside crystalline solids. Also, the existence of hidden or dark eigenwaves is predicted, which may exhibit topologically nontrivial properties. 
As a practical example, we investigate this atomistic photonic eigenwaves in a zincblende semiconductor SiC, which has been widely studied for multiple applications, such as CMOS compatible single-photon quantum sources~\cite{castelletto2014,widmann2015,nagy2019}. 
Our study of the microscopic optical responses inside semiconductors is crucial for light-emitting applications at the nanoscale and emerging light-based quantum technologies.

\section{Deep microscopic optical band structure and eigenwaves}
In materials science, electron band structure (Fig.~\ref{fig:2}c) represents the allowed energy of electronic states of a crystal within the Brillouin zone (Fig.~\ref{fig:2}a and \ref{fig:2}b); along with electron wave functions, this band theory characterizes the electronic properties of the crystal. Likewise, a band theory for phonon (Fig.~\ref{fig:2}d) defines the thermal and mechanical properties of the crystal. These quantities are at the heart of materials science that establishes the fundamental understanding of a crystalline solid. In this section, we will develop the deep microscopic optical band structure (Fig.~\ref{fig:2}e) and deep microscopic optical waves which completely characterize the optical responses of a crystal~\cite{bharadwaj2024}.

Here, we examine the band theory for optical responses in a crystalline lattice. The Fourier representation of the inverse dielectric matrix in the reciprocal lattice space relates the external and total longitudinal electric fields as 
    \begin{equation}
        E_{}^L(\mathbf{q+g},\omega)=\sum_{\mathbf{g}'}\varepsilon_{\mathbf{gg'}}^{-1}(\mathbf{q},\omega)E_0^L(\mathbf{q+g'},\omega),
    \end{equation}
where $E^L$ and $E_0^L$ are the total and external fields, respectively, $\varepsilon_{\mathbf{gg'}}$ are the elements of dielectric matrix, \textbf{g} and \textbf{g}’ are the reciprocal lattice vectors, \textbf{q} is the crystal-wavevector within the Brillouin zone, and $\omega$ is the frequency. Here, we note that the dielectric matrix contains all possible optical responses of the crystal including the symmetry information of the crystal~\cite{po2017}, and can be expressed as the symmetry-indicating lattice conductivity $\sigma_{\mathbf{gg'}}(\mathbf{q},\omega)$. However, those information is hidden behind the complexity of dielectric matrix. We obtain the microscopic optical band structure and the hidden atomistic polarization waves of a material through eigen-decomposition of the response matrix as~\cite{bharadwaj2024}
    \begin{equation}
        \bar{\varepsilon}(\mathbf{q},\omega)=\sum_{\lambda}{|E_\lambda\rangle \pi_\lambda \langle E_\lambda|},
    \end{equation}
where $\bar{\varepsilon}$ denotes the dielectric matrix, $\lambda$ is the eigenmode index, $\pi_\lambda(\mathbf{q},\omega)$ and $|E_\lambda(\mathbf{q},\omega)\rangle$ are the corresponding microscopic optical index and eigenwave, respectively. Equivalently, the eigen-decomposition is expressed as\cite{lu2008} $\pi_{\lambda}^{-1}(\mathbf{q},\omega)E_{\lambda}^L(\mathbf{q+g},\omega)=\sum_{\mathbf{g}'}\varepsilon_{\mathbf{gg'}}^{-1}(\mathbf{q},\omega)E_\lambda^L(\mathbf{q+g'},\omega)$,
    where $E_\lambda^L(\mathbf{q+g},\omega)$ is the Fourier component of the eigenwave. This microscopic optical index contains the symmetry information of the crystal, so the deep microscopic optical eigenwave is a symmetry-indicated quantity~\cite{bharadwaj2024}. 
Here, the eigenvalue of inverse dielectric matrix $\pi_\lambda^{-1}(\mathbf{q},\omega)$ is related to the screening strength $1-\pi_\lambda^{-1}$ that the corresponding eigenwave experiences in the crystal, and eigenwaves distinguish the fields with different screening strengths. In the limiting case of $1-\pi_\lambda^{-1} \rightarrow 0$, the total field equals the external field, indicating that no screening occurs for the corresponding eigenwave.

The dielectric matrix is Hermitian at the frequency below the bandgap, so the microscopic optical eigenwaves are obtained as a set of orthonormal basis functions. Therefore, the eigenwaves can represent the optical waves inside the crystalline lattice. For longitudinal fields, the optical waves exist in the presence of external driving sources, e.g., an electron beam. A self-sustained longitudinal (charge-density oscillation) wave is obtained at the poles of the dielectric responses as plasmon resonances. In terms of the microscopic optical eigenwaves, plasmon resonances occur when $\pi_\lambda(\mathbf{q},\omega)=0$. The plasmon resonance spectra can be described in terms of the microscopic optical eigenwaves as the energy loss function $-\mathrm{Im}[\pi_\lambda(\mathbf{q},\omega)^{-1}]$, analogous to the commonly used expression of $-\mathrm{Im}[\varepsilon(\mathbf{q},\omega)^{-1}]$ for electron energy loss spectra. This representation will be useful for probing energy loss spectra at the high momentum regime~\cite{poursoti2022}.

Analogous to the dark modes widely discussed in nanophotonics community, \textit{hidden eigenwaves} can be defined as the microscopic optical eigenwaves that do not interact with a macroscopic planewave. In fact, our analysis shows that there exist many dark or hidden eigenwaves that we discuss in the next section. We expect these hidden eigenwaves may have unconventional properties including topological optical $N$-invariant~\cite{van2021,van2022}, and excitation of such hidden eigenwaves will be investigated in the future work.

    \begin{figure}
        \centering
        \includegraphics[scale=0.5]{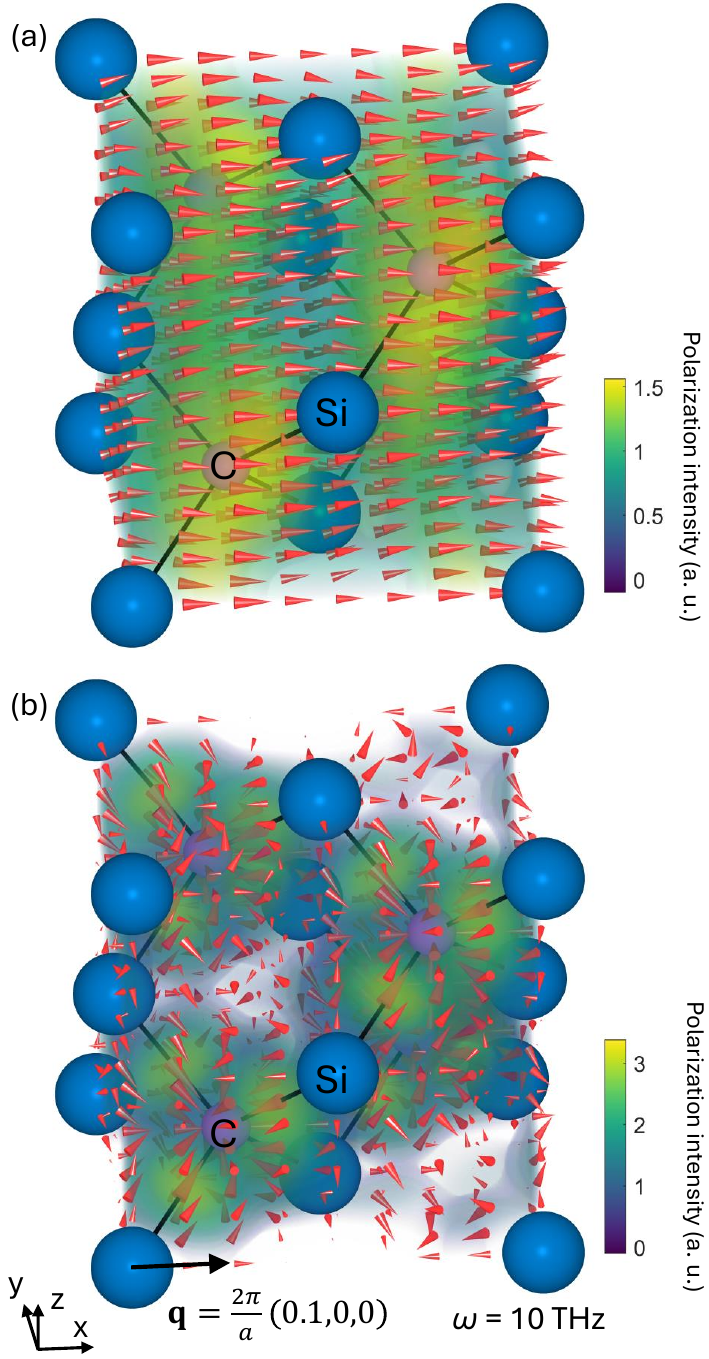}
        \caption{Distributions of atomistic lights and local polarization textures in a crystal. The volumetric intensity distributions of microscopic optical eigenwaves (colored volume) and their polarization vectors (red arrows) are visualized for (a) the dominant band ($\lambda=0$) with nearly planewave feature and (b) a higher-order band ($\lambda=1)$ with rapidly-varying polarization textures at the lattice level. Silicon and carbon atoms are indicated by blue and purple spheres, respectively. The polarization vectors are visualized at a certain time frame; the temporal evolution of the polarization texture over the optical cycle is provided in the Supplementary Movies~\cite{si}. The crystal momentum is $\mathbf{q}=\frac{2\pi}{a}(0.1,0,0)$, and the frequency is $\omega$ = 10 THz.}
    \label{fig:dist}
    \end{figure}

\section{Real-space representation of deep microscopic optical waves}
By analyzing the real-space distribution of eigenwaves, the microscopic optical responses of a material, which have been hidden behind macroscopic homogenization, can be visualized. The real-space representation of longitudinal eigenwaves is $\mathbf{E}_{\lambda\mathbf{q}}^L(\mathbf{r},\omega)=\sum_{\mathbf{g}}E_\lambda^L(\mathbf{q+g},\omega)\hat{n}_\mathbf{q+g}e^{i(\mathbf{q+g})\cdot\mathbf{r}}$, where the longitudinal unit vector is $\hat{n}_\mathbf{q+g}=\frac{\mathbf{q+g}}{|\mathbf{q+g}|}$, and the induced charge density corresponding to a longitudinal eigenwave can be obtained using the Poisson equation. The induced charge density represents how the charge density is deformed at the corresponding eigenwave fluctuation and shows somewhat remnant forms of the tetrahedral bonds of SiC (Fig.~S2). Upon the perturbations, the electronic states oscillate between the ground and excited states, and the induced charge density visualizes this virtual transition happening inside the crystal.

The dominant microscopic optical eigenwave ($\lambda=0$) shows nearly planewave feature (Fig.~\ref{fig:dist}a) at a finite crystal momentum of $\mathbf{q}=\frac{2\pi}{a}(0.1,0,0)$ and frequency of 10~THz. This mode dominantly represents the classical macroscopic planewave near the optical limit. However, higher-order microscopic optical eigenwaves exhibit rapidly-varying polarization textures at the lattice level (Fig.~\ref{fig:dist}b). A higher-order band with $\lambda=1$ shows polarization vectors localized at the silicon-carbon bonds and pointing inwards and outwards with respect to the carbon atoms (See Supplementary Movie 2 for the temporal evolution of this mode~\cite{si}). This mode visually shows polarization singularity around the carbon atom and potentially represents a hidden eigenwave with topologically nontrivial optical properties.

\begin{figure}[]
        \centering
        \includegraphics[scale=0.5]{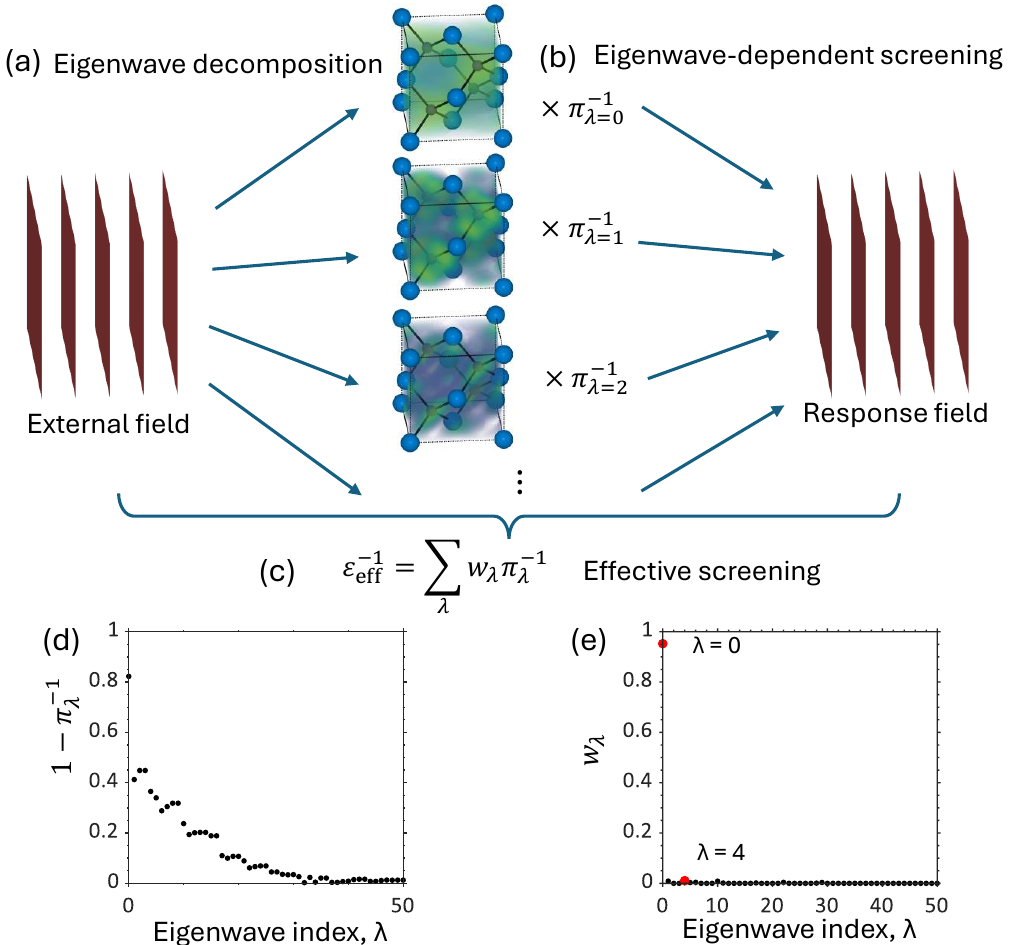}
        \caption{Schematics of the effective screening process experienced by an arbitrary field in a crystal and the evolution of macroscopic dielectric constant. (a) An external field is decomposed into a set of eigenwaves, where the weight is determined by the overlap between the external field and the eigenwaves. (b) Each eigenwave experiences different screening strengths in the solid. (c) The response field experiences an effective screening strength as a weighted average of the screening strengths of the eigenwaves. The macroscopic dielectric constant is obtained by using a macroscopic planewave as the external field. (d) The screening strength, $1-\pi_{\lambda}^{-1}$, of eigenwaves at the momentum $\mathbf{q} = \frac{2\pi}{a}(0.1,0,0)$. (e) The eigenmode weight, $w_\lambda$, for macroscopic planewave. Only the eigenwaves with the mode indices $\lambda$ = 0 and 4 contribute more than 1\% to the macroscopic dielectric constant with $w_0=0.9525$ and $w_4=0.0107$. The frequency is $\omega$ = 10~THz.}
    \label{fig:screening}
    \end{figure}

Another interesting result of our work is the existence of nonplanar inhomogeneous microscopic optical eigenwaves even at the optical limit near the $\Gamma$-point. These higher-order eigenwaves with $\lambda>0$ exhibit rapidly varying distributions (Fig.~S2), implying that nonclassical atomistic responses may occur even at the small momentum limit. The eigenwaves with $\lambda=2$ and $3$ have degenerate screening strength, and their distributions are related by four-fold improper rotation along the $x$-axis, as expected from a zincblende structure (Fig.~S2a and S2b). The eigenwave with $\lambda=4$ exhibit the polarization direction parallel to the crystal momentum (Fig.~S2d), which allows it to weakly couple with a longitudinal macroscopic planewave (Fig.~\ref{fig:screening}e). However, other higher-order eigenwaves have large mode mismatch with the macroscopic planewave, which restricts their contribution to the classical refractive index.

The eigenwaves of a single band exhibit significantly different behavior along different symmetry points. The dominant band ($\lambda=0$) near the $\Gamma$-point represents the classical optical limit $|\mathbf{q}|\rightarrow 0$, and the corresponding eigenwave distribution resembles a planewave. The polarization distribution is nearly homogeneous and in‑phase, compared to other eigenwaves. Away from the $\Gamma$-point, the eigenwaves exhibit more localized polarization textures. Such rapidly varying distribution at the subwavelength level indicates the optical responses at the atomistic lattice level associated with the local-field effect, or the momentum exchange processes involving the optical diffraction due to the atomic lattice. At the X-point, phase variation along the propagation exists, and an out-of-phase response can be observed from the inverted color of the induced charge density (Fig.~S1b). Similarly, at the L-point, an out-of-phase response along the (111) direction is observed (Fig.~S1c). 

\section{Effective dynamic screening and macroscopic dielectric constant}
Now, we will examine the effective screening process experienced by an arbitrary field in a crystal. Above, we obtained a set of orthonormal eigenwaves with distinct screening strengths using a Hermitian dielectric matrix. These eigenwaves can be used to decompose a general field and obtain the effective screening strength that the field experiences. The effective screening strength can be obtained as $\varepsilon_{\mathrm{eff}}^{-1}=\sum_\lambda w_\lambda \pi_\lambda^{-1}$, where the weight is determined by the overlap between the external field and the eigenwaves as $w_\lambda=\sum_{\mathbf{g}}|E_{\lambda\mathbf{g}}^\dagger E_\mathbf{g}|^2$ (Fig. \ref{fig:screening}a). Therefore, a field experiences an effective screening strength in a crystal (Fig. \ref{fig:screening}c), which leads to the effective dielectric constant. How the effective screening strength evolves is determined by the eigenwaves and their corresponding screening strengths, and how the external field couples with the eigenwaves.

We may examine the macroscopic dielectric constant using this picture. For a macroscopic planewave, the weight is given as $w_\lambda=\sum_{\mathbf{g}}|E_{\lambda\mathbf{g}}^\dagger \delta_{\mathbf{g}0}|^2$. Although there are many eigenwaves with nonnegligible screening strengths (Fig.~\ref{fig:screening}d), only a few of them contribute to the macroscopic dielectric constant (Fig.~\ref{fig:screening}e). By examining their mode distribution, we can easily see that the eigenwaves that strongly couple with the macroscopic planewave have large mode matching (Fig.~\ref{fig:dist}a). Most eigenwaves have negligible mode matching with the macroscopic planewave, so they are not excited by or couple with the planewave and have small or no contribution to the macroscopic dielectric constant (Fig.~\ref{fig:dist}b). Another interesting observation is that the macroscopic dielectric constant obtained in this way, $\varepsilon_\mathrm{eff}=4.908$, is smaller than the macroscopic dielectric constant obtained without the local-field effect, $\varepsilon_{00}=5.435$. It can be explained from the reduced effective screening strength due to the involvement of higher-order eigenwave ($\lambda=4$) with lower screening strength. This observation agrees with the literature where the macroscopic dielectric constant is lowered by considering the local-field effects.

The refractive index of a matter is not a uniquely determined quantity, but depends on the structure of light inside the media. The screening strength decreases quite rapidly for the higher-order eigenwaves (Fig.~\ref{fig:screening}d). If we consider a structured light that couples strongly with such high-order eigenwaves, then its effective screening strength will be much lower than that of macroscopic planewaves. This effect will be especially prominent for point sources and localized light, which have large higher Fourier components~\cite{baumberg2019,gonccalves2020}. Nano and picoscale optics with small feature sizes and localized light will be subject to this lowered effective dielectric constant.

This work shows the role of microscopic optical band structure and eigenwaves as a fundamental quantity to derive optical properties of solids. Further investigations are expected on the inclusion of the transverse parts of eigenwaves for general optical interactions~\cite{sangalli2017,bharadwaj2024}, the characterization of nonlinear responses using the eigenwaves~\cite{sipe2000}, and maximally localized Wannier function representation of the eigenwaves~\cite{marzari2012}.

\section{Conclusions}
In conclusion, we unravel microscopic optical band structure of SiC. The microscopic optical band provides quantum generalization of refractive index and reveals hidden polarization waves. 
The microscopic optical eigenwaves has distinct screening strengths in the crystal, and the effective screening experienced by an arbitrary field is obtained as an average of these screening strengths, where the weight is obtained by the coupling between the field and eigenwaves. As a direct consequence, the classical refractive index is not a unique quantity even for an identical crystal, but also depends on the interacting field. This proposed theoretical framework provides a solid physical picture of the dielectric constant and elevates the microscopic optical eigenwave as a fundamental quantity representing the optical properties of matter. Also, hidden optical eigenwaves are predicted in a crystalline solid that are not excited by a macroscopic planewave.
 In the future, this framework will be extended to two-dimensional materials (e.g., graphene and MoS$_2$) and transverse excitation.

\section*{Acknowledgments}
This work was supported by the Office of Naval Research (ONR) under the award number N00014231270. J.M. acknowledges the NRF \textit{Sejong} Science fellowship funded by the MSIT of the Korean government (RS-2023-00252778).


\appendix

\bibliography{bibliography.bib}

\end{document}



\title[]{Supplementary information for\\Visualization of atomistic optical waves in crystals}
\author{Jungho Mun}
    \affiliation{Electrical and Computer Engineering, Purdue University, West Lafayette, Indiana 47907, USA}
    \affiliation{POSCO-POSTECH-RIST Convergence Research Center for Flat Optics and Metaphotonics, Pohang University of Science and Technology (POSTECH),\\ Pohang 37673, Republic of Korea}
\author{Sathwik Bharadwaj}
    \affiliation{Electrical and Computer Engineering, Purdue University, West Lafayette, Indiana 47907, USA}
\author{Zubin Jacob}
    \email{zjacob@purdue.edu}
    \affiliation{Electrical and Computer Engineering, Purdue University, West Lafayette, Indiana 47907, USA}
    \affiliation{Purdue Quantum Science and Engineering Institute, Birck Nanotechnology Center, Purdue University, West Lafayette, Indiana 47907, USA}

\maketitle

\section{Calculation of the microscopic optical band structure and eigenwaves}
    The elements of dielectric matrix were evaluated using the random-phase approximation and first-order perturbation theory according to Adler-Wiser theory~\cite{adler1962,wiser1963,bharadwaj2022} as
    \begin{equation}
        \varepsilon_{\mathbf{gg'}}(\mathbf{q},\omega) = \delta_{\mathbf{gg'}}-v_{\mathbf{gg'}}(\mathbf{q})
        \sum_{\mathbf{k}ll'} (f_{\mathbf{k+q},l'}-f_{\mathbf{k}l})
        \frac{
        \bra{\psi_{\mathbf{k}l}}e^{-i(\mathbf{q+g})\cdot\mathbf{r}}\ket{\psi_{\mathbf{k+q}l'}}
        \bra{\psi_{\mathbf{k+q}l'}}e^{i(\mathbf{q+g'}_n)\cdot\mathbf{r}}\ket{\psi_{\mathbf{k}l}}
        }{\mathcal{E}_{\mathbf{k+q},l'}-\mathcal{E}_{\mathbf{k}l}+\hbar\omega^+},
    \label{eqn:eps_LL}
    \end{equation}
    where 
    $\psi_{\mathbf{k}l}$ is the electronic Bloch function, 
    $l$ is the band index,
    $\mathcal{E}_{\mathbf{k}l}$ is the energy for the electronic band,
    $f_{\mathbf{k}l}$ is the Fermi-Dirac distribution function,
    $\mathbf{g}$ is the reciprocal lattice vector,
    and $v_{\mathbf{gg'}}(\mathbf{q})=\frac{4\pi e^2}{|\mathbf{q+g}||\mathbf{q+g'}|}$ is the symmetrical Coulomb kernel for making the dielectric matrix hermitian, which is different from ref.~\cite{adler1962,wiser1963}.
    In the dielectric matrix calculation, we employed an energy cutoff 400~eV for planewave grids and a cutoff 80~eV for the dielectric matrix, which corresponds to the size of the matrix of $51\times51$. Electron wave functions were obtained using the empirical pseudopotential method, where 4 valence bands and 20 conduction bands were considered. The Brillouine zone integration was performed using the $k$-point mesh of $15\times15\times15$. The planewave grids and $k$-point mesh grids were chosen to satisfy the point group symmetries of face-centered-cubic lattice. The microscopic optical band structure and eigenwaves are directly obtained from the calculated dielectric matrix using the eigen-decomposition as
    \begin{equation}
        \pi_\lambda(\mathbf{q},\omega)E^L_\lambda(\mathbf{q+g},\omega)=\sum_{\mathbf{g}'}\varepsilon_{\mathbf{gg'}}E^L_\lambda(\mathbf{q+g'},\omega),
    \end{equation}
    where $\lambda$ is the eigenmode index, $\pi_\lambda(\mathbf{q},\omega)$ is the microscopic optical index or the eigenvalue, and $E^L_\lambda(\mathbf{q+g},\omega)$ is the Fourier components of the microscopic optical eigenwave or the elements of eigenvector. The real-space representation of microscopic optical eigenwave is
    \begin{equation}
        \mathbf{E}^L_{\mathbf{q}\lambda}(\mathbf{r},\omega)=\sum_{\mathbf{g}}E^L_\lambda(\mathbf{q+g},\omega)\hat{n}_\mathbf{q+g}e^{i\mathbf{(q+g)\cdot r}},
    \end{equation}
    where the longitudinal unit vector is $\hat{n}=\frac{\mathbf{q+g}}{|\mathbf{q+g}|}$. The induced charge density associated with the eigenwave is obtained as
    \begin{equation}
        \delta\rho_{\mathbf{q}\lambda}(\mathbf{r},\omega)=-\frac{1}{4\pi i}\sum_\mathbf{g}E^L_\lambda(\mathbf{q+g},\omega)|\mathbf{q+g}|e^{i\mathbf{(q+g)\cdot r}}.
    \end{equation}
    
    \texttt{Purdue-PicoMax}. The numerical calculations were performed using our in-house code, which will be released as an open-source software package \texttt{Purdue-PicoMax} in the future. The development of electronic structure calculators based on the density functional theory has significantly extended our understanding of electronic and phononic properties of materials. \texttt{Purdue-PicoMax} will be released for various semiconducting materials including Moire materials, and we expect this software will contribute to picoscale optical materials engineering. 

\begin{figure*}
    \centering
    \includegraphics[scale=0.7]{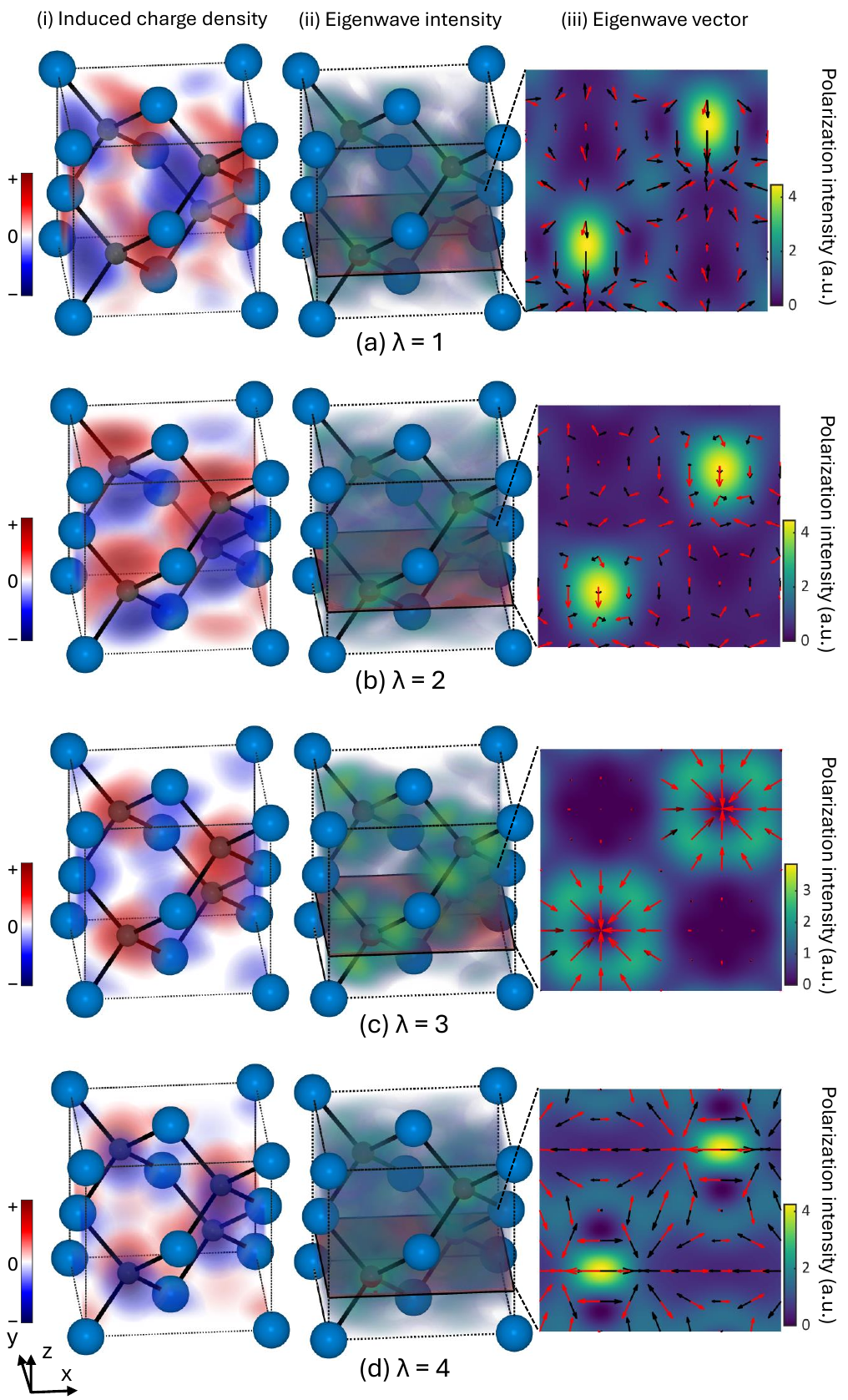}
    \caption{Distributions of higher-order microscopic optical eigenwaves ($\lambda=1,2,3,4$) at the $\Gamma$-point. (i) Induced charge density distribution, where red and blue color indicates the positive and negative values. (ii) Volumetric intensity distribution of microscopic optical eigenwaves, (iii) and their polarization vectors (arrows) in the cross-sectional views (red plane).}
    \label{fig:s2}
\end{figure*}

\begin{figure*}
    \centering
    \includegraphics[scale=0.7]{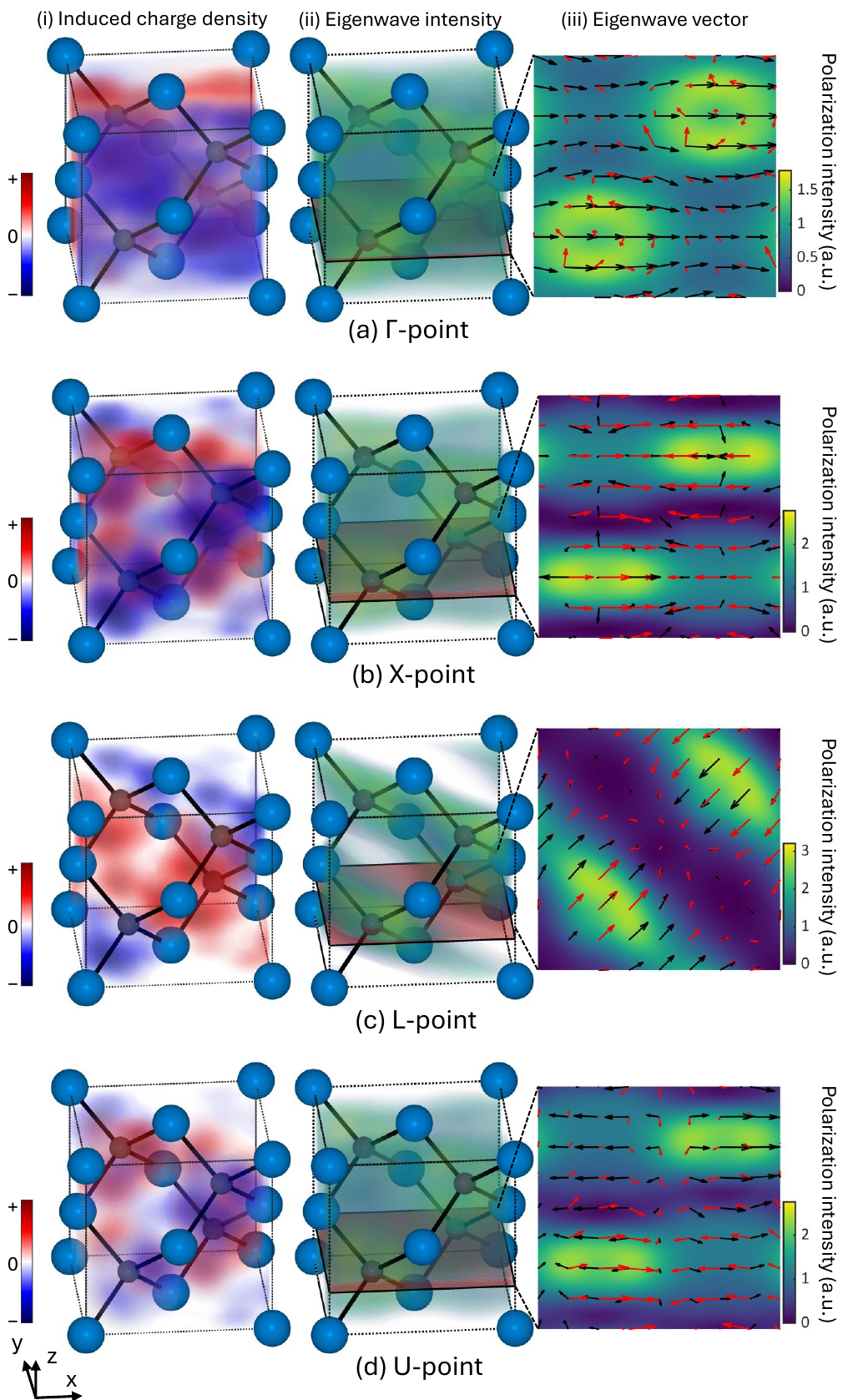}
    \caption{Distributions of the dominant microscopic optical eigenwaves ($\lambda=0$) at some high-symmetry points. (i) Induced charge density distribution, where red and blue color indicates the positive and negative values. (ii) Volumetric intensity distribution of microscopic optical eigenwaves, (iii) and their polarization vectors (arrows) in the cross-sectional views (red plane). (a) $\Gamma$-point, (b) X-point, (c) L-point, and (d) U-point.}
    \label{fig:s1}
\end{figure*}

\clearpage
\section*{Supplementary Movie Captions}
\textbf{Supplementary Movie 1.} The temporal evolution of the dominant microscopic optical eigenwave ($\lambda=0$) over one optical cycle. The volumetric color indicates the intensity of the eigenwave (polarization intensity), and the red arrows indicate the eigenwave vector (polarization texture). This eigenwave corresponds to Fig.~3a and b of the main text; the crystal momentum is $\mathbf{q}=\frac{2\pi}{a}(0.1,0,0)$ and the frequency is $\omega$ = 10 THz.

\textbf{Supplementary Movie 2.} The temporal evolution of a higher-order microscopic optical eigenwave with $\lambda=1$ over one optical cycle. The volumetric color indicates the intensity of the eigenwave (polarization intensity), and the red arrows indicate the eigenwave vector (polarization texture). This eigenwave corresponds to Fig.~3c and d of the main text; the crystal momentum is $\mathbf{q}=\frac{2\pi}{a}(0.1,0,0)$ and the frequency is $\omega$ = 10 THz.

\bibliography{bibliography.bib}